# Animating the development of *Social Networks* over time using a dynamic extension of multidimensional scaling

Loet Leydesdorff,[a] Thomas Schank,[b] Andrea Scharnhorst,[c] & Wouter De Nooy[d]

[a] Amsterdam School of Communication Research (ASCoR), University of Amsterdam, Kloveniersburgwal 48, 1012 CX Amsterdam, The Netherlands; loet@leydesdorff.net ; http://www.leydesdorff.net (*corresponding author*).
[b] Technical University of Karlsruhe, Faculty of Informatics, ITI Wagner, Box 6980, 76128 Karlsruhe, Germany; schank@ira.uka.de.
[c] The Virtual Knowledge Studio for the Humanities and Social Sciences (VKS), Royal Netherlands Academy of Arts and Sciences (KNAW), Cruquiusweg 31, 1019 AT Amsterdam, The Netherlands; andrea.scharnhorst@vks.knaw.nl.
[d] Amsterdam School of Communication Research (ASCoR), University of Amsterdam, Kloveniersburgwal 48, 1012 CX Amsterdam, The Netherlands; w.denooy@uva.nl .

## Abstract

The animation of network visualizations poses technical and theoretical challenges. Rather stable patterns are required before the mental map enables a user to make inferences over time. In order to enhance stability, we developed an extension of stress-minimization with developments over time. This dynamic layouter is no longer based on linear interpolation between independent static visualizations, but change over time is used as a parameter in the optimization. Because of our focus on structural change versus stability the attention is shifted from the relational graph to the latent eigenvectors of matrices. The approach is illustrated with animations for the journal citation environments of *Social Networks*, the (co-)author networks in the carrying community of this journal, and the topical development using relations among its title words. Our results are also compared with animations based on *PajekToSVGAnim* and *SoNIA*.

*Corresponding author*: Loet Leydesdorff, Amsterdam School of Communication Research (ASCoR), University of Amsterdam, Kloveniersburgwal 48, 1012 CX Amsterdam, The Netherlands.
Tel: +31-20-5256598; Fax: +31-8-42239111; email: loet@leydesdorff.net

# Animating the development of *Social Networks* over time using a dynamic extension of multidimensional scaling

## Abstract

The animation of network visualizations poses technical and theoretical challenges. Rather stable patterns are required before the mental map enables a user to make inferences over time. In order to enhance stability, we developed an extension of stress-minimization with developments over time. This dynamic layouter is no longer based on linear interpolation between independent static visualizations, but change over time is used as a parameter in the optimization. Because of our focus on structural change versus stability the attention is shifted from the relational graph to the latent eigenvectors of matrices. The approach is illustrated with animations for the journal citation environments of *Social Networks*, the (co-)author networks in the carrying community of this journal, and the topical development using relations among its title words. Our results are also compared with animations based on *PajekToSVGAnim* and *SoNIA*.

**Keywords:** animation, network, dynamic, stress, structure, evolution

## Introduction

When one extends the visualization of networks at each moment in time to the animation of these networks over time, one encounters a number of technical and theoretical problems. The technical ones involve the stability of the representation because the human mind needs to be able to entertain a mental map on the basis of the animation (Misue *et al*., 1995; Moody *et al*., 2005; Bender-deMoll & McFarland, 2006). The pattern in the representation has to be stabilized while computer layouts can be mapped, rotated, and sized independently from one instant to another, using criteria like stress-minimization. The consequent series may be distorted or discontinuous to such an extent that one would no longer be able to draw a mental map or to infer hypotheses about network growth or structural changes from the resulting animations.

More fundamentally, an adjacency matrix of a network contains only relational information, not coordinates of the nodes in a metric space. In a static representation of a network positions can be deduced from the network of relations. In a dynamic animation, however, the researcher may precisely be interested in changes in positions of nodes. This would require relative stability in positioning the nodes (and not the links). Boerner *et al*. (2005), for example, solved this problem by using the last map in a time series to fix the positions of the nodes, and backtracked from this map to its evolution over time by erasing nodes and links from the perspective of hindsight, while keeping the positions constant.

The visualization program *Pajek* (De Nooy *et al*., 2005)[1] allows for a strategy in which one first generates a configuration using the aggregate of networks over time as a baseline. Given this initial layout, one can then use the positions at each time $t$ as a starting point for the optimization at time $t + 1$. The thus generated sequence can be saved and used as input to *PajekToSVGAnim.exe*.[2] This program generates an SVG-animation which can be brought online. Using their program *Social Networks Image*

[1] *Pajek* is freely available for non-commercial usage at http://vlado.fmf.uni-lj.si/pub/networks/pajek/.
[2] *PajekToSVGAnim.Exe* is available at http://vlado.fmf.uni-lj.si/pub/networks/pajek/SVGanim/default.htm.

*Animator* (*SoNIA*),[3] Moody *et al.* (2005) define time windows which span sets of relational events. The time windows ("bins") can be overlapping or not. In these animations, the nodes move as functions of relations (Bender-deMoll *et al*., 2006). The animations of *SoNIA* can be exported as QuickTime movies.

In this study, we take an approach different from the interpolation between independent snapshots at different moments in time. We submit a dynamic extension of social network analysis based on stress minimization and multidimensional scaling (MDS). Stress can be minimized both at each instant in time and over time (Baur & Schank, 2008). This algorithmic approach of the problem of relating static and dynamic representations was recently implemented in *Visone*, another publicly available program for the visualization of social network data. A version of this program with the dynamic extension is available online at http://www.leydesdorff.net/visone/index.htm.

The potential of our approach is illustrated below with animations of: (1) the position of the journal *Social Networks* among other journals in its citation impact environment (1994-2006), (2) the development of the coauthor networks in *Social Networks* during the period 1988-2007, (3) the topical network in terms of co-occurrences of title words, and (4) the knowledge base of the publications in *Social Networks* in terms of their aggregate references to other journals. Furthermore, the pros and cons of using this new algorithm for the animation are discussed in relation to *PajekToSVGAnim.Exe* and *SoNIA* as two (non-commercial) alternatives.

---

[3] *SoNIA* is available at http://www.stanford.edu/group/sonia/.

## Theoretical and methodological considerations

When two or more visualizations are generated for different moments in time, one is intuitively inclined to attribute the observable change to changes in the system under study. However, one should be aware that the positions of nodes and links in the visualization of a network are due to algorithms which optimize the visualization using different criteria for the layout, for example, in order to avoid the unnecessary crossing of edges. Any map remains a projection in two dimensions of a multi-dimensional object. Thus, changes in the visualization over time can be attributed to developments in the system to be visualized or differences in the optimizations and/or the angle of the projection.

If one considers the relations as variables developing over time and in relation to one another, one would have to consider the partial differential equations for all these variables. This would lead to an analytically almost never solvable problem. At issue is that a hidden variable may cause dependency relations among the observable variables. If one is interested in evolving structures, for example, eigenvectors of the matrices (underlying the networks) can be considered as such hidden variables. Indeed, the hypothesis of eigenvectors is based on assuming spurious correlations among the observables. If both the factor loadings and the factors themselves are allowed to vary over time, the models become unidentifiable without further assumptions.

Latent dimensions can be reconstructed from the observable data, for example, by using factor analysis or MDS. (In our applications, we sometimes use one-mode data, e.g., co-authorship relations, but in most cases we use two-mode data, e.g., words as variables *versus* documents as cases.) For the purpose of multivariate analysis the (rectangular) attribute matrix or, in graph-theoretical terms, the two-mode network (Wasserman & Faust, 1994, at pp. 29-30, 154) must first be transformed into a (square) matrix of similarities or dissimilarities. A pattern of shared attributes can be transformed into a similarity score, e.g., a Pearson correlation coefficient (Wasserman & Faust, 1994, at p. 386). If the similarity matrix is based on the symmetrical adjacency matrix, however, the results can no longer be expected to reflect the (eigen-)structure in the data (Leydesdorff & Vaughan, 2006).[4]

In summary, positional information can be extracted from, e.g., a Pearson correlation matrix which is generated on the basis of an attribute matrix. The extraction can be done algorithmically using factor analysis (MDS, principal component analysis, correspondence analysis,[5] etc.) and/or by human pattern recognition on the basis of the information contained in a similarity matrix (or a visualization thereof). Using MDS, one previously had to pencil groupings and relations into the resulting visualizations of positions (Leydesdorff, 1986; Leydesdorff & Cozzens, 1993). Network visualization programs, however, have the advantage above traditional MDS that the links, partitions, and clusters can be visualized by the software.

[4] The one-mode adjacency matrix or sociomatrix contains less information than the two-mode attribute matrix because it is generated by multiplying the latter with its transposed. However, it is not possible to generate an attribute matrix from the adjacency matrix.

[5] In the case of (quasi-)correspondence analysis, the similarity measure is implied by the use of chi-square statistics (Faust, 2005; De Nooy, 2003).

Ahlgren *et al.* (2003) argued that the Pearson correlation is inferior to the cosine for the purpose of showing similarities in the case of sparse (attribute) matrices. Normalization to the mean (as in the case of a Pearson correlation) can then be counterproductive. Technically, the cosine is equal to the Pearson correlation coefficient, but without the normalization to the mean (Jones & Furnas, 1987; Leydesdorff & Zaal, 1988).[6] Using the cosine matrix as input to the visualization, one visualizes a vector space (Salton & McGill, 1983). The vector space has a topology different from the relational space since it represents coordinates. Distances in it are positional measures of similarity in the distributions of relations. In this study, we shall reflexively use both topologies.

### Stress minimization

The projection of a multi-dimensional object into fewer dimensions requires the minimization of stress in the projection. Network visualization programs use algorithms for this which differ from traditional MDS.

Kruskal & Wish (1978) defined the stress value for MDS as follows (cf. Borgatti, 1998):

---

[6] The cosine is formulated as follows:

$$\text{Cosine}(x,y) = \frac{\sum_{i=1}^{n} x_i y_i}{\sqrt{\sum_{i=1}^{n} x_i^2}\sqrt{\sum_{i=1}^{n} y_i^2}} = \frac{\sum_{i=1}^{n} x_i y_i}{\sqrt{(\sum_{i=1}^{n} x_i^2) * (\sum_{i=1}^{n} y_i^2)}} \quad (1)$$

where $x_i$ and $y_i$ refer to the score of the $i^{th}$ row (e.g., document) in column $x$ or $y$ (e.g., different words).

$$S = \sqrt{\frac{\sum_{i \neq j} (\|x_i - x_j\| - d_{ij})^2}{\sum_{i \neq j} d_{ij}^2}} \tag{2}$$

In terms of network analysis, $\| x_i - x_j \|$ is the actual distance in the layout between each pair of nodes *i* and *j*, whereas the parameter $d_{ij}$ in this formula represents the graph-theoretical distance making the shortest path between these two nodes. Note that *i* and *j* now refer to different columns (or rows) in a symmetric similarity matrix.

In their seminal work, Kamada & Kawai (1989) reformulated the problem of achieving graph-theoretical target distances in terms of energy optimization. They formulated the ensuing stress in the graphical representation as follows:

$$S = \sum_{i \neq j} \frac{1}{d_{ij}^2} (\|x_i - x_j\| - d_{ij})^2 \tag{3}$$

Equation 3 differs from Equation 2 not only because of the square root, but more importantly because of the weighting of each term with $1/d_{ij}^2$ in Equation 3. This weight is crucial for the quality of the layout, but defies normalization with $\sum d_{ij}^2$ in the denominator (as in Equation 2). In other words, the two stress values cannot be compared.

Kamada & Kawai (1989) used a gradient descent method to iteratively minimize the stress according to Equation 3. Gansner *et al.* (2004) improved on Kamada & Kawai's algorithm by minimizing the majorant of *S*. This function can be minimized efficiently by using matrix methods. In a number of empirical case studies, these authors showed that their approach leads to faster convergence, is less sensitive to local minima, and improves on the remaining stress. Furthermore, the minimization of the majorant can be implemented using an algorithm that is more compact than that of Kamada & Kawai (1989). The minimization is performed locally and, therefore, can more easily be modified. We use this opportunity below for solving the problem of how to relate static and dynamic layouts.

The localized method moves a node *i* in dimension *d* to its new location according to:

$$\text{new-}x_i^{(d)} = \frac{\sum_{j \neq i} w_{ij} \left( x_j^{(d)} + d_{ij} \frac{x_i^{(d)} - x_j^{(d)}}{\| x_i - x_j \|} \right)}{\sum_{j \neq i} w_{ij}} \qquad (4)$$

Accordingly, the iterative algorithm can be formalized as follows:

```
repeat
      for each node i
            in each dimension d
                  x_i^(d) ← new- x_i^(d)
until the stress is minimized appropriately.
```

**Table 1**: Semi-code for the local minimization of stress.

The outer loop can be repeated for a fixed number of steps or until the stress no longer improves more than a small fraction. If one assumes a fixed number of steps, this algorithm runs in $\theta(n^2)$ time.

One of us extended this algorithm in order to layout dynamic networks. The dynamic stress function is in this case provided by the following equation:

$$S = \left[ \sum_t \sum_{i \neq j} \frac{1}{d_{ij,t}^{\;2}} (\|x_{i,t} - x_{j,t}\| - d_{ij,t})^2 \right] + \left[ \sum_{1 \leq t < |T|} \sum_i \omega \, \|x_{i,t} - x_{i,t+1}\|^2 \right] \qquad (5)$$

In Equation 5, the left-hand term is equal to the static stress in Equation 3, while the right-hand term adds the dynamic component, namely the stress between subsequent years. If the weighting factor $\omega$ for this dynamic extension is set equal to zero, the method is equivalent to the static analysis, and the layout of each time frame is optimized independently. The dynamic extension penalizes drastic movements of the position of node $i$ at time $t$ ($x_{i,t}$) toward its next position ($x_{i,t+1}$) by increasing the stress value. Thus, stability is provided in order to preserve the mental map between consecutive layouts so that an observer can identify corresponding graph structures.

In other words, the configuration for each year can be optimized in terms of the stress in relation to the solutions for previous years and in anticipation of the solutions for following years. In principle, the algorithm allows us (and *Visone* enables us) to extend this to more than a single year, but in this study the optimization is extended by only one year in both directions (that is, including $t + 1$ and $t - 1$). Note that this approach is different from the approach that takes the solution for the previous moment in time as a starting position for iterative optimization along recursive trajectories. The nodes are not repositioned given a previous configuration, but the entire previous and next configurations are included in the algorithmic analysis for each year.

Technically, the equation to be optimized in each iteration computes a new position for each node ($x_i$) on dimension $d$ as follows:

$$\text{new-}x_{i,t}^{(d)} = \frac{\left[\sum_{j \neq i} w_{ij,t} \left( x_{j,t}^{(d)} + d_{ij,t} \frac{x_{i,t}^{(d)} - x_{j,t}^{(d)}}{\left\| x_{i,t} - x_{j,t} \right\|} \right) \right] + \omega \ (x_{i,t-1}^{(d)} + x_{i,t+1}^{(d)})}{\left[\sum_{j \neq i} w_{ij,t}\right] + 2\omega} \tag{6}$$

until the aggregated stress falls below a threshold value or during a fixed number of iterations. Again, the left-hand terms (between brackets) in both the numerator and the denominator of Equation 6 account for the static solution, while the right-hand terms contain the extensions with the stress in comparison to the previous ($t$–1) and next ($t$+1) moments in time. Higher values of the weighting factor for the dynamic extension ($\omega$) result in increased stability of the representations over the years.

The corresponding algorithm can now be written as follows:

```
repeat
      for each time t
            for each node i
                  in each dimension d
```

$$x_{i,t}^{(d)} \leftarrow \texttt{new-}\, x_{i,t}^{(d)}$$

```
until the stress is minimized appropriately.
```

**Table 2**: Semi-code for the dynamic extension of stress minimization.

Under the same assumption as above—that is, a fixed number of steps—the algorithm runs in $\theta(n^2 \cdot |T|)$ time. Therefore, the proposed algorithm for dynamic layouts does not impose an asymptotic running time larger than computing the static layouts for each moment in time separately.

**Methods and materials**

In order to demonstrate the new algorithm, we apply it to a number of scientometric analyses of the journal *Social Networks.* The relevant publication and citation data were harvested from the Web-of-Science edition of the *Social Science Citation Index* for the period 1988-2007 on January 27, 2008. The document set of 425 titles corresponds to the Volumes 10 to 29 of the journal.[7] This data was used to construct two-mode matrices (for each of the years) of documents as cases versus authors, title words, and cited references, respectively, as variables. (Dedicated software routines for these purposes are made available by one of us at http://www.leydesdorff.net/software.htm.)

[7] The download includes 10 more documents of the third and fourth issues of 1987 which appeared only in 1988 and were therefore included in the download, but not in this analysis.

Words listed as stopwords at http://www.uspto.gov/patft/help/stopword.htm were excluded from the co-word analysis. Single occurrences of variables in each year were also deleted. 165 title words occurred more than once in a single year. Among the 445 authors publishing in *Social Networks* during this period, 364 coauthored with one another in this context. 146 of these co-authors were part of the largest component in any of the years under study, and therefore included in the analysis.

Using the *Journal Citations Reports* of the (*Social*) *Science Citation Index*, aggregated journal-journal citation data is additionally available in electronic format since 1994 and till 2006. Dedicated routines were used to construct aggregated journal-journal citation matrices from this data using *Social Networks* as the seed journal. All journals with a citation relation to *Social Networks* were included in the analysis, in the 'cited' and 'citing' dimension, respectively. (This leads to two different citation matrices.) Note that an aggregated journal-journal citation matrix is asymmetrical (cited *versus* citing), and therefore two structures can be analyzed. We shall first focus on the citation impact environment of *Social Networks*, that is, all journals citing articles from *Social Networks* (Leydesdorff, 2007a).

Similarity matrices (using the cosine as similarity criterion) were constructed on the basis of the column vectors of the two-mode matrices. Because the cosine varies between zero and one, a threshold has to be set.[8] In most cases, we shall use cosine ≥ 0.2 as (a relatively arbitrary) threshold value.[9] Nodes which do not attach to the aggregated component(s) in any of the years under study were removed in order to reduce noise in the animations.

[8] When one uses the Pearson correlation matrix, one can set the threshold at $r \geq 0$ because $r$ varies between -1 and +1. However, there is no one-to-one relation between the cosine and the Pearson correlation coefficient (Egghe & Leydesdorff, in preparation).

### Further methodological choices in the dynamic layouter

One of the "Layout" options among the menus of *Visone* is called "Dynamic." Within this menu, one can select various parameters such as the stability parameter $\omega$. In the animations of this study, we chose unity for this parameter, 120% for the length of links, and 0.6 for the component separation. Increasing the length of links in computing standard graph-theoretical distances can help to improve on the visualization in dense regions of the network.

The initial layout for the aggregated set is computed in *Visone* by using *Pivot-MDS* (Brandes & Pich, 2007), followed by static stress minimization (Equation 2 above). An advantage of this procedure is that MDS provides a deterministic basis for the animations so that the resulting configurations can always be reproduced. The method of Kamada & Kawai (1989) and its derivations—such as minimization of the majorant in our next steps—do not work on disconnected graphs without modification. If $G_t$ is disconnected some of the target distances $d_{ij,t}$ can become infinite during the distance computation.

---

[9] In both *SoNIA* and *Visone* one can use the cosine-matrix without a threshold for spanning the vector space, and thereafter use a threshold value only for the visualization. However, this option is not available in Pajek (De Nooy *et al.*, 2005), and we used Pajek in this study for the pre-processing (see below).

In the current implementation, these values are substituted by the maximum non-infinite distance in $G_t$ (that is, $d_{\max,t} = \max_{i \neq j} \left\{ d_{ij,t} : d_{ij,t} \neq \infty \right\}$ ) or the maximum of all these distances ( $d_{\max} = \max_{t} \left\{ d_{\max,t} \right\}$ ). In this study, we chose the latter option because whatever can be kept constant over the years may add to the stability in the representation. However, we found that both these options tend to separate components more than desired. In order to ameliorate this $d_{max,t}$ or $d_{max}$ can optionally be scaled by a factor between zero and one. As noted, we chose 0.6 for this parameter.

*Visone* allows for the attribution of centrality values to the nodes which can be used for coloring or sizing the nodes in the animation. In the dynamic version, endurance of the nodes over the years can additionally be visualized as shades of colors. Links (edges and arcs) can be qualified both in width and color according to their respective values. In this study, animations were screen-captured and recorded using BlueBerry's Flashback ™. This program allows for editing of the animations and export into the flash format which is ready for upload on the Internet.

**Results**

*a. aggregated journal-journal citation structures*

In the early 1980s several research teams more or less at the same time and independently realized that aggregated journal-journal citation listings, as provided by the Journal

Citations Reports (JCR) of the *(Social) Science Citation Index* could be used for the generation of networks which exhibit structure in scientific communication (Doreian & Farraro, 1985; Leydesdorff, 1986; Tijssen *et al*., 1987). The JCR data has been available in electronic format since 1994.[10] The time series contains the information required to study the changing patterns of citations among journals, for example, in the case of interdisciplinary or emerging developments in the sciences. These aggregated developments are beyond control of individual agents and, therefore, provide us with a potential baseline for measuring the effects of (e.g., governmental) interventions and priority programming (Studer & Chubin, 1980, at pp. 269 ff.; Leydesdorff, 1986; Leydesdorff *et al*., 1994; Leydesdorff & Schank, forthcoming).

The journal *Social Networks* was established in 1978. The corresponding field of network analysis became topical among scholars in other disciplines in the aftermath of the emergence of the Internet during the 1990s (Scharnhorst, 2003). Algorithms developed by social network analysts such as various forms of centrality (Freeman, 1978/1979; Wasserman & Faust, 1994) became core concepts in the newly emerging network sciences. Did this change the citation impact of *Social Networks* itself? Did the journal become a hub in an increasingly interdisciplinary network among social scientists, applied mathematicians, and physicists interested in the distributional properties of the Internet?

[10] The Web-of-Science version is available since 1998.

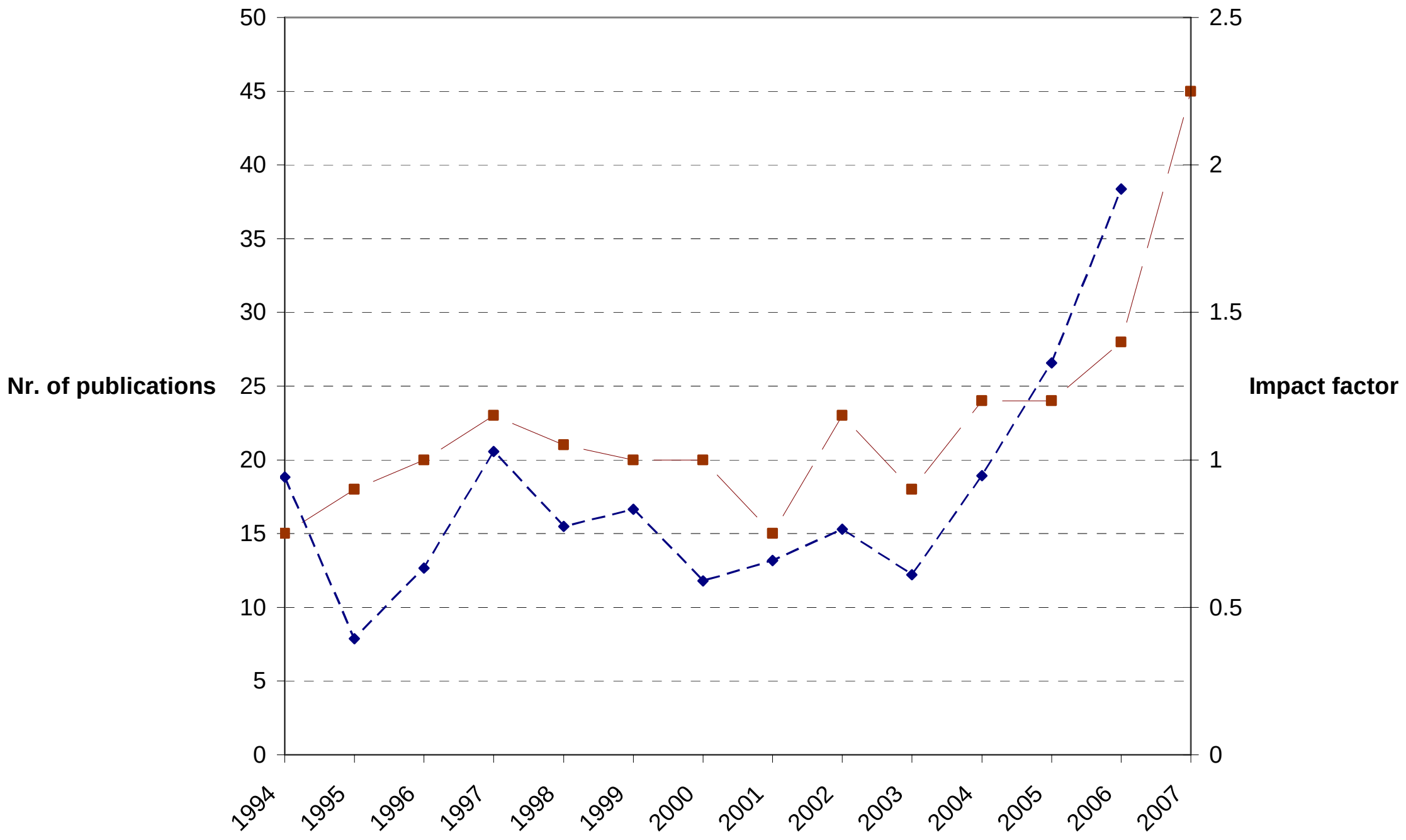

**Figure 1**: Number of publications (■) and impact factors (♦), *Social Networks* 1994-2007.

Figure 1 shows the development of the number of publications and the impact factor of *Social Networks* since 1994. Both the size and the impact of the journal have increased steadily since 2003. However, the impact factor is a global measure that, while it may vary dramatically among fields of science, does not inform us about where the journal has impact in terms of fields of science (Leydesdorff, 2008a).

Leydesdorff (2007a) suggested the use of local citation environments of specific journals instead of global impact factors, and brought these environments online as cosine matrices for all journals in the ISI-set at http://www.leydesdorff.net/jcr06. Leydesdorff (2007b) added that betweenness centrality in the vector space might be used as an indicator of interdisciplinarity of a journal, and illustrated this with Figure 2 for the citation impact environment of *Social Networks*.

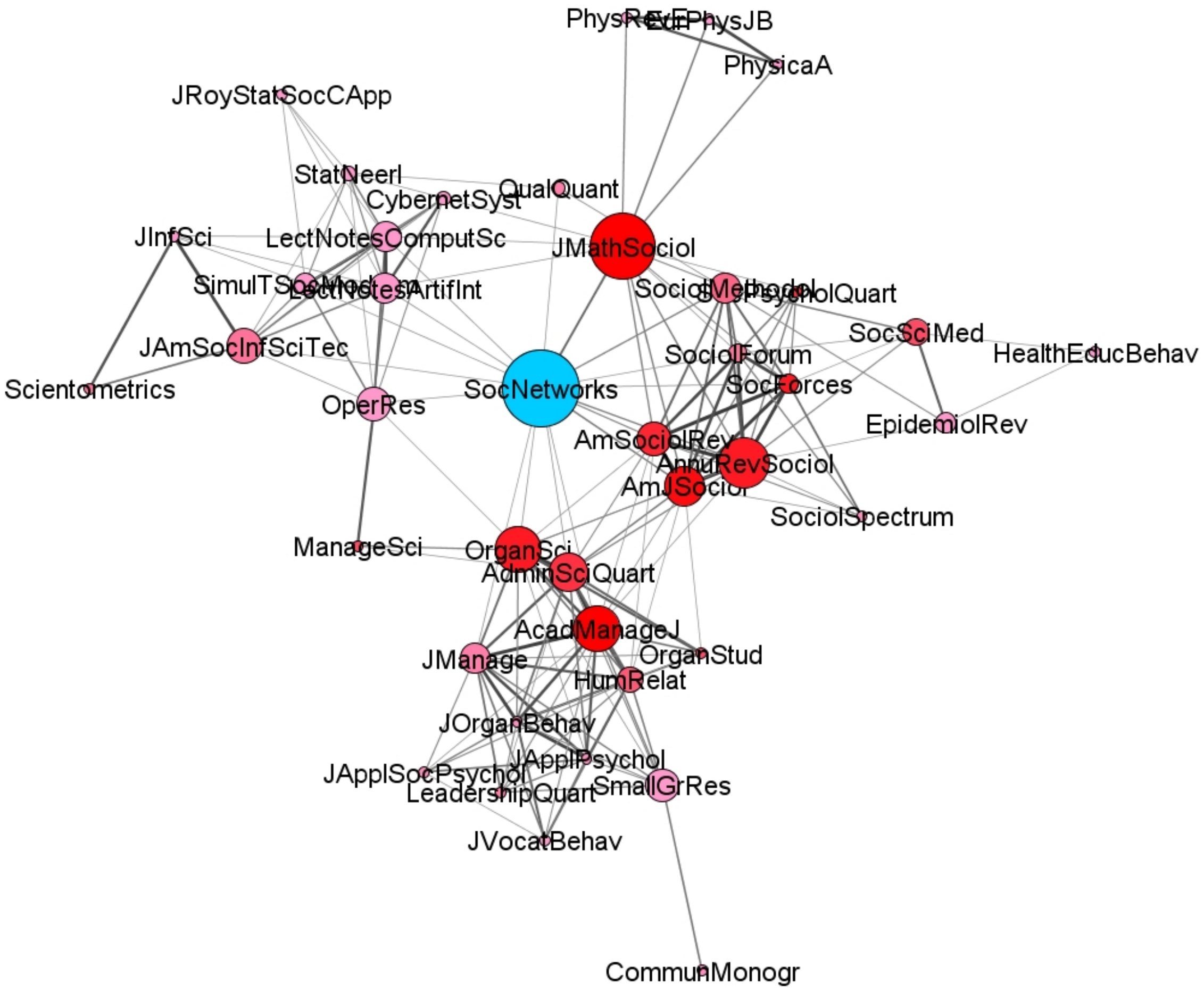

**Figure 2**: The local citation impact environment of *Social Networks* in 2004. (Sizes of nodes are proportional to betweenness centrality in the vector space; cosine ≥ 0.2.)

Figure 2 provides the local citation impact[11] environment of *Social Networks* 2004 in the format that we shall use in the animations. The sizes of the nodes correspond to the betweenness centrality and the color shades to endurance in this citation environment. In 2004, *Social Networks* functioned as a bridge between two social science clusters (sociology and organization studies), a computer science cluster including some statistics journals, and, related via the *Journal of Mathematical Sociology,* a physics cluster. The time series, however, will inform us that this interdisciplinary position was exceptional rather than the rule.

[11] The choice of the word 'impact' is to be considered technically. A reference can be expected to mean different things in different contexts (MacRoberts & MacRoberts, 1987; Leydesdorff & Amsterdamska, 1990).

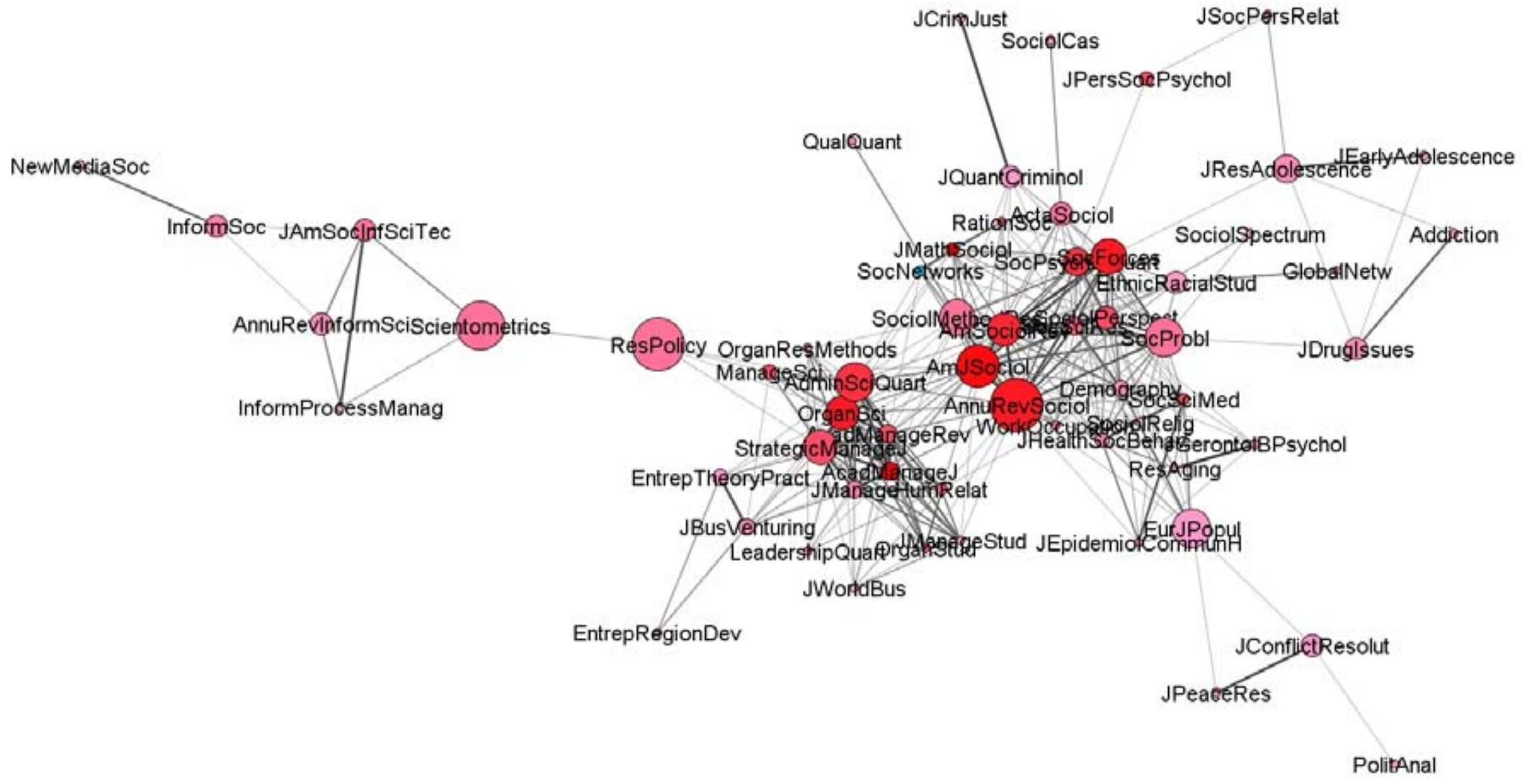

**Figure 3**: 63 journals in the local citation environment of *Social Networks* in 2006.

Figure 3 shows the local citation environment of *Social Networks* in 2006. In this year, papers in *Social Networks* were cited by papers in 78 journals of which 63 were related to the main component with a cosine value larger than or equal to 0.2. Two major clusters are involved in most of the recent years: one among journals in sociology and another among journals in organization and management studies. *Social Networks* itself is part of the former cluster.

The animation at http://www.leydesdorff.net/journals/socnetw/index.htm teaches us that during the period under study, *Social Networks* has been part of the cluster of sociology journals in most of the years, while journals in organization and management studies increasingly became part of its citation impact environment. In some years, the citation impact of *Social Networks* reaches beyond these two sets. However, this position cannot be sustained as shown in Figure 4 which reveals the development of the betweenness centrality of *Social Networks* during these years.[12] In the terminology of nonlinear dynamics and chaos theory, one could say that the journal makes excursions from its basin-of-attraction in some of the years, but thereafter returns to its disciplinary position.

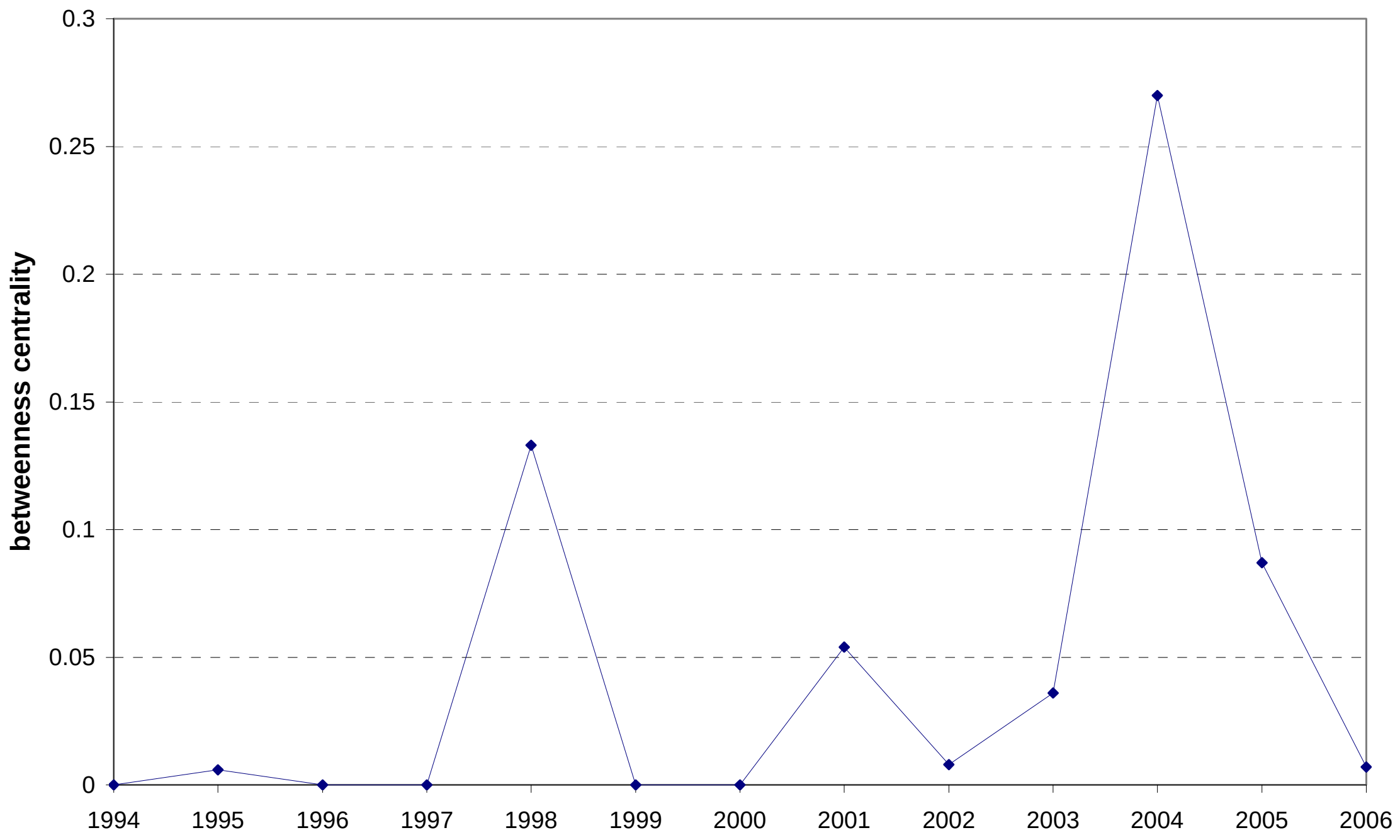

**Figure 4:** Betweenness centrality of *Social Networks* in its citation impact environment.

[12] The development of this betweenness centrality and the impact factor are not correlated ($r = 0.07$; *n.s.*)

In summary, *Social Networks* is primarily part of a set of journals in sociology. In most years, it is well embedded in this group of journals which relates *as a group* with journals in management and organization studies. In some years, *Social Networks* is cited in a larger citation environment including journals in physics and applied mathematics, but this is an exception rather than the rule. In spite of the fact that the citation impact of *Social Networks* in recent years has increased, this has not changed its disciplinary identity.

The transposed matrices provide us with the citing structures of *Social Networks.* In this case, the relevant citation environment consists of all journals from which articles are cited in *Social Networks* in a particular year. The corresponding animation can be retrieved from http://www.leydesdorff.net/journals/socnetw/citing.htm. While the citation impact environment provides us with a visualization of the relevant environment of *Social Networks* as a source, this animation provides us with an impression of the citation behavior of authors within the journal: how do these authors reconstruct their field in terms of relevant references?

In terms of cited references made by authors within the journal, *Social Networks* is embedded in a sociology set of journals even more firmly than in the cited dimension. Journals in social psychology provide a more continuous source of references than organization and management studies, although in more recent years the latter have become increasingly important. The relation with social psychology in the (re)construction of this field is understandable given the important role of this discipline in the genesis of social network analysis (Freeman, 2004).

In summary, the journal cannot be considered as an interdisciplinary journal in its contribution to the reconstruction of aggregated journal-journal relations; it is rather a specialist journal with citation impacts outside sociology as a discipline. In the terminology of Gould & Fernandez (1990), *Social Networks* can be considered as a representative of sociology journals.

*b. the coauthor network among publications in Social Networks*

For the coauthorship analysis we use a dedicated routine (available at http://www.leydesdorff.net/software/coauth/index.htm) for each respective volume of *Social Networks* during the period 1988-2007. As noted, this routine provides us in each year with the (two-mode) attribute matrix, the cosine-normalized matrix, and the adjacency matrix. In the case of co-authorship analysis, we focus on the adjacency matrices for the animation because coauthors form a relational network. The animation is brought online at http://www.leydesdorff.net/journals/socnetw/coauth/index.htm.

As noted above, the 425 documents in volumes 10 to 29 of *Social Networks* were written by 445 unique authors of whom 146 coauthored during the period under study as participants of a main component. Figure 5 first shows the aggregated MDS of this component (Brandes & Pich, 2007).

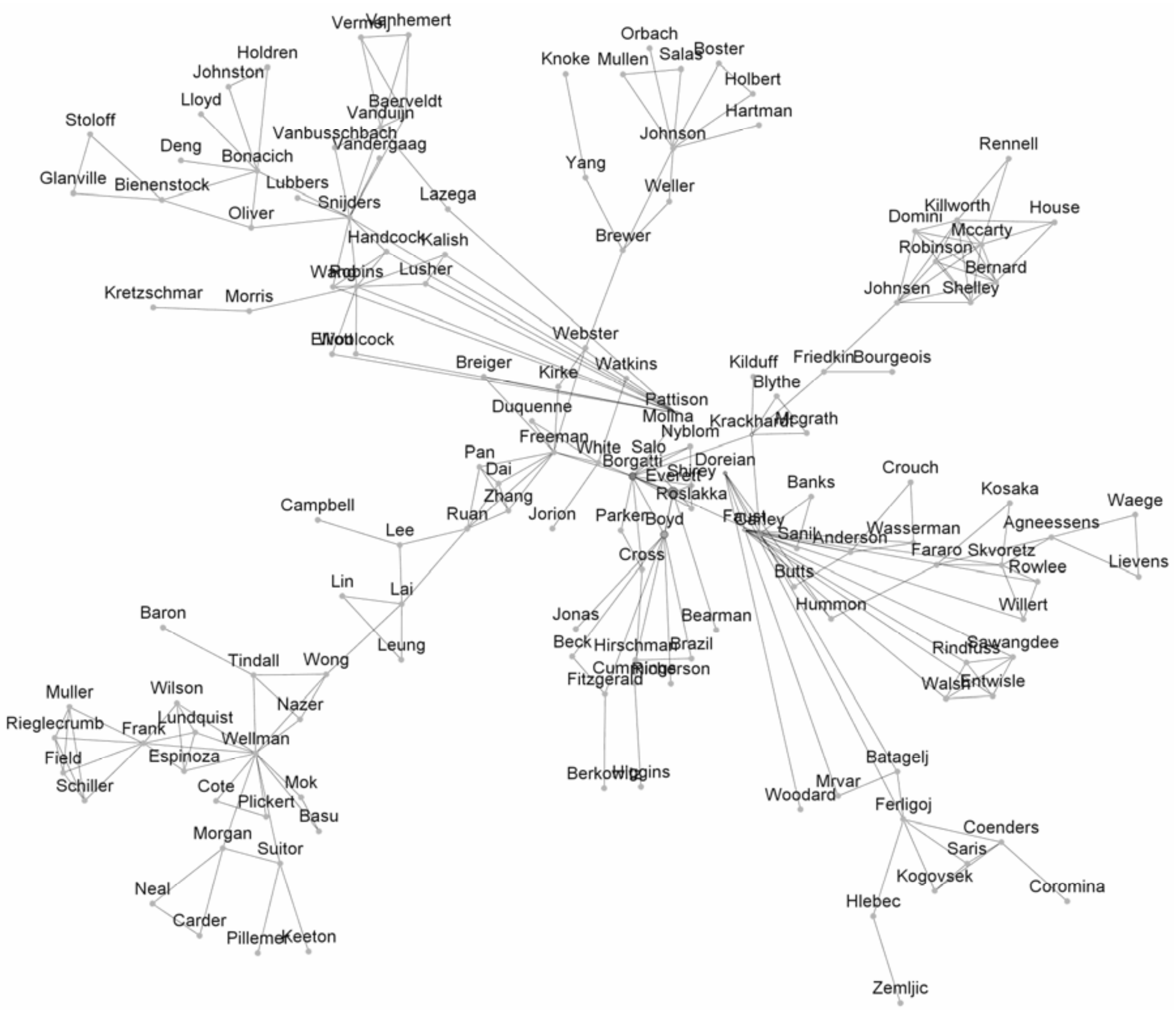

**Figure 5**: MDS map of the main component of 146 coauthoring authors in *Social Networks* 1988-2007.

This graph visualizes the carrying community of the journal and specialty. At the top left, one finds scholars developing new statistical methods for social network analysis. These are Exponential Random Graph Models analyzed by Mark Handcock in Washington, Garry Robins in Melbourne, and coworkers (Robins & Morris, 2007; Robins *et al*., 2004, 2007a, and 2007b), statistical estimation of longitudinal network data analyzed by a group around Tom Snijders in Groningen and Oxford (Baerveldt & Snijders, 1994; Van Duijn *et al.,* 1999; Lubbers & Snijders 2007; Snijders 1990; Van Der Gaag & Snijders, 2005), and Bonacich's continuous work with coworkers on (eigenvector) centrality (Bonacich 1991, 2007; Bonacich *et al*., 1998, 2004; Bonacich & Lloyd, 2001, 2004).

The interest in social ties centered around Barry Wellman and his NetLab at the University of Toronto (Mok & Wellman 2007; Plickert *et al*. 2007; Suitor *et al*., 1997; Wellman 1996, 2007; Wellman *et al.,* 1991, 1997) is represented by a Canadian dominated cluster at the bottom left. The bottom right features the Slovenian group around Vladimir Batagelj and Anuška Ferligoj working on Pajek software and algorithms (Batagelj & Mrvar, 2000, 2001), blockmodels with Patrick Doreian (Batagelj, 1997; Batagelj *et al*., 1992a and b; Doreian *et al*., 2004a and b), and measurement validity (Coromina *et al*., 2008; Ferligoj & Hlebec 1999; Hlebec & Ferligoj, 2001; Kogovsek & Ferligoj, 2005; Kogovsek *et al*., 2002).

At the middle right, one finds the first scholars who worked on statistical models for (random) networks including Stanley Wasserman (Anderson *et al*., 1992, 1999a; Wasserman & Anderson, 1987), John Skvoretz and Thomas J. Fararo (Fararo & Skvoretz, 1984; Fararo *et al*., 1994; Skvoretz, 1982, 1985, 1990, 1991; Skvoretz *et al*., 2004). They are linked to other parts of the network by other pioneers of social network analysis, notably Katherine Faust (Anderson *et al*., 1992; Faust 1988, 1997; Faust *et al*., 2000, 2002; Faust & Wasserman, 1992) and Kathleen M. Carley (Anderson *et al*., 1999b; Borgatti *et al*., 2006; Carley & Krackhardt, 1996; Hummon & Carley, 1993; Sanil *et al*., 1995).

At the top right, we retrieve the Florida-based group around H. Russell Bernard , Peter D. Killworth and Christopher McCarty (Bernard *et al*., 1990; Johnsen *et al*., 1995; Killworth *et al*., 1998; Killworth *et al*., 2003, 2006; McCarty *et al*., 1997; Shelley *et al*., 1995) specializing in data collection.[13] In the center, we find scholars who have contributed to social network analysis on diverse fronts for a long time, including Linton C. Freeman (Freeman, 1996; Freeman *et al*., 1991, 1998; Freeman & Duquenne 1993; Keul & Freeman, 1987; Ruan *et al*., 1997), Ronald Breiger (Breiger, 2005; Breiger & Pattison, 1986; Pattison & Breiger, 2002), Martin G. Everett and Stephen P. Borgatti (Borgatti & Everett 1989, 1992a, 1992b, 1993, 1994, 1997, 2000, 2006; Everett & Borgatti, 1988, 1990, 1993, 1996, 2000, 2005), Philippa Pattison (Lazega & Pattison, 1999; Pattison, 1988; Pattison & Breiger, 2002; Robins *et al*., 2001, 2004, 2007a and b), David Krackhardt (Borgatti *et al*., 2006; Carley & Krackhardt, 1996; Friedkin & Krackhardt, 2002; Krackhardt 1988; Krackhardt & Kilduff, 2002; McGrath *et al*., 1997), and Douglas R. White (Freeman *et al*., 1991; White & Duquenne, 1996; White, 1996; White & Borgatti, 1994; White & Jorion, 1996).[14] For several of these authors *Social Networks* has been a primary outlet for publications and some of them are or have been (associate) editors of this journal.

In summary, the multidimensional scaling of the aggregated coauthorship network groups the founding fathers and mothers of social network analysis as a methodology with branches representing more specialized developments. Some of these developments are rather recent, such as the new statistical models and generalized blockmodeling, while other specialties are almost as old as social network analysis itself such as issues of network data collection and the analysis of random graphs.

[13] The top (middle) cluster is an artifact resulting from the grouping of three different authors with the name Johnson. We intend to correct the software so that first initials are taken into account. (Next initials may generate error again because authors are not always using these next initials.) In this case, however, two of these three authors would have had a "J" as their first initial.
[14] Kevin White coauthoring with Susan Watkins in an article in 2000 is not Douglas R. White who is involved in the other coauthorship ties in *Social Networks*.

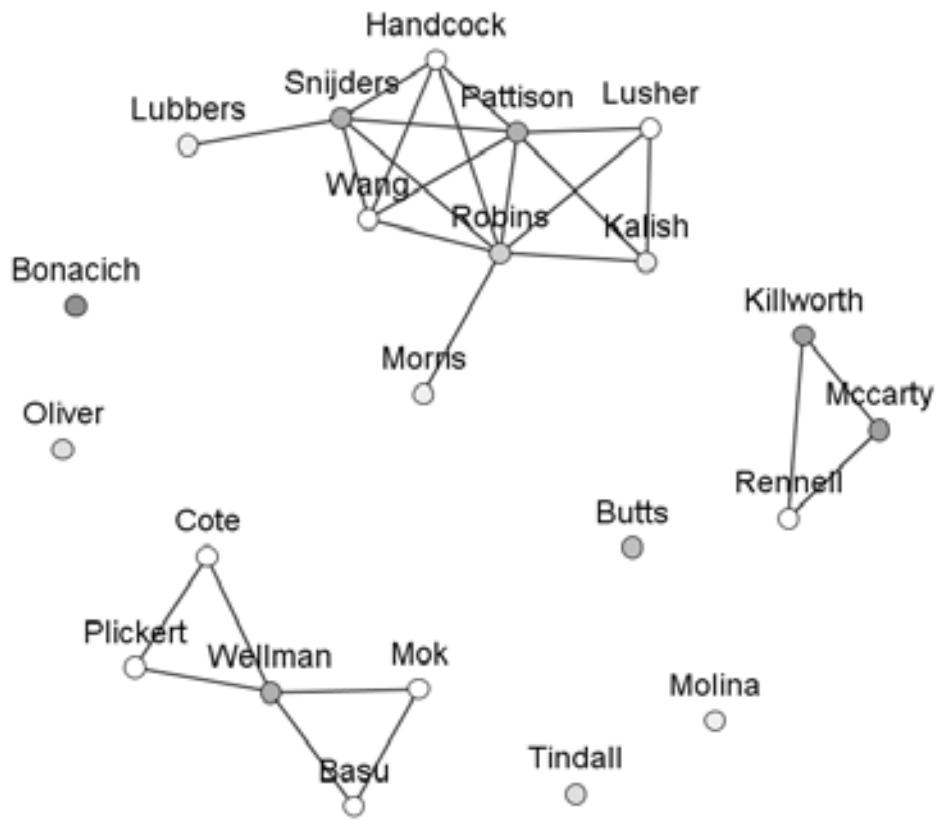

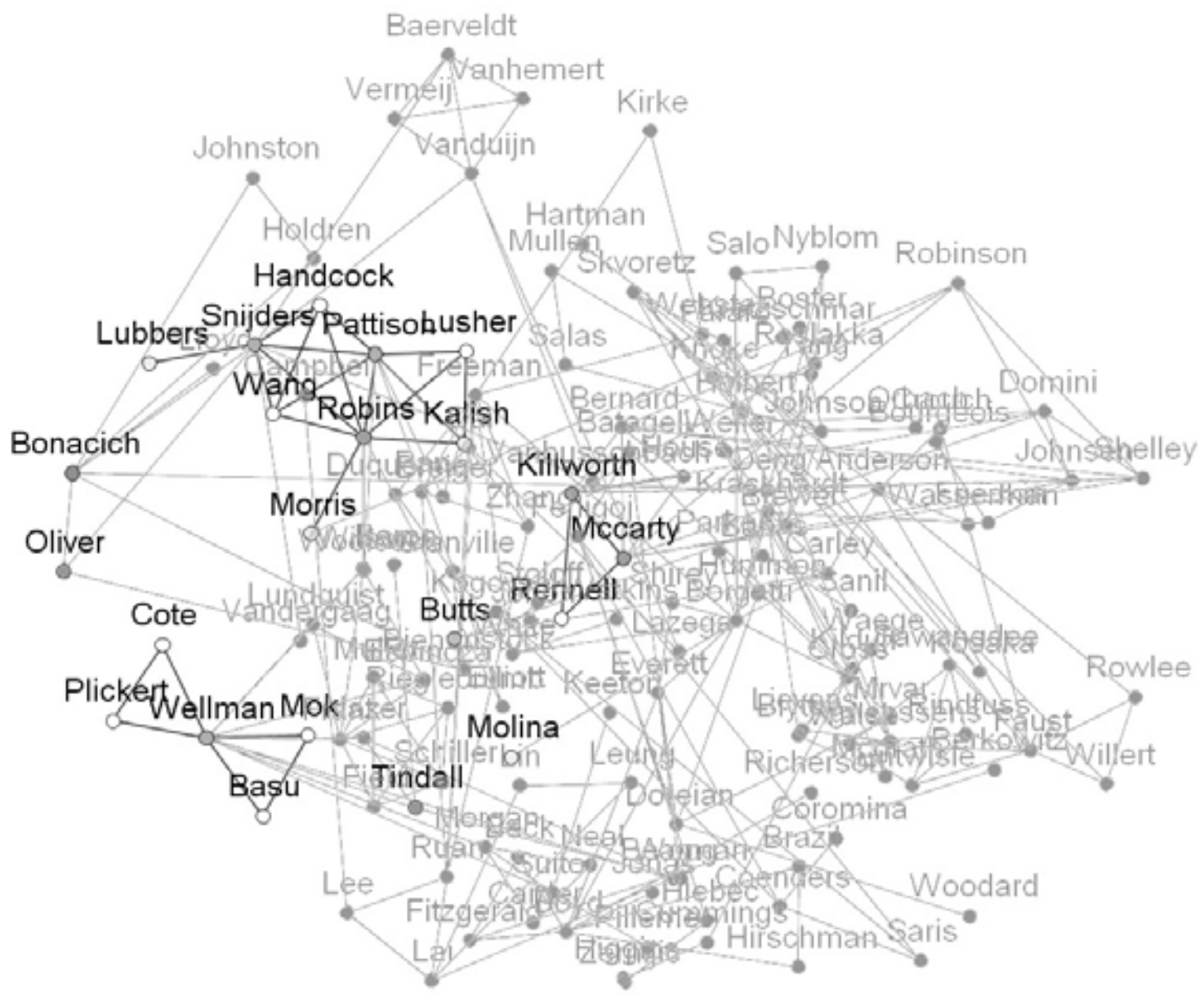

**Figure 6a:** The author and coauthor map of *Social Networks* 2007.

**Figure 6b**: The 2007 map of Fig. 6a superimposed on the aggregated network 1988-2007.

Figure 6a shows the configuration in 2007. Twenty-two authors belonging to the large component are authoring or co-authoring in this year. (The five single-authored presences are related to the main component because of co-authorship relations in other years.) In summary, coauthorship relations in a specific year provide us only with instantiations of the network which is formed by this community. The observable structures of each year show how the relations vary, given a background of communal interests.

In Figure 6b, the map of 2007 (Fig. 6a) is superimposed on the aggregated network for the same year. In the animation—available online at http://www.leydesdorff.net/journals/socnetw/coauth/au_aggr.htm—this aggregate of relations over the years is added as a background. The animation begins with the full picture of the configuration in 1988 (Figure 5 above). In subsequent years, various authors from the periphery are drawn into the circle by coauthorship relations. A core group is thus shaped whose members, however, are not visible in terms of single-authored or coauthored publications in visualizations for each year separately.

If we focus on the movements of authors within the center, that is, watch the movements of authors that are already in the center of the network, we see that Linton C. Freeman is moving around a lot in the first part of this period, signifying that he is collaborating with people working in diverse specialties. This also applies to Douglas R. White. It seems apt to conceive of them as pioneering generalists. In the later part of the period, Stephen P. Borgatti, and to a lesser extent Kathleen M. Carley, are the most volatile nodes. They seem to represent generalists of the second generation, combining work on network-analytic methodologies in general with a specialization in substantive applications.[15]

In summary, the sparseness of the co-author networks and the clarity of the aggregated collaboration network, suggest that the mental map is improved by animating the accumulative network (retaining all previous collaborations as changes to the structure of relations). This collective construction is not visible on a year to year basis, but a structure among the central authors is reproduced in terms of authorship and coauthorship by a carrying community. As the community of (co)authors is constructed over the years, their network is increasingly contracted towards a center.

[15] Transition to the cosine-normalized matrices (available at http://www.leydesdorff.net/journals/socnetw/coauth/cosine.htm) does not change the nodes or their relations when compared with the animation based on the adjacency matrices. In this case, there are no large differences among the players (as there is among journals) and therefore normalization does not make much difference.

*c. Title words*

During the period 1988-2006, 165 title words occurred more than once in a single year, and were included in the analysis (after the correction for stopwords). An animation of this is provided at http://www.leydesdorff.net/journals/socnetw/ti/index.htm. Through the course of time, we see particular issues reappear, notably centrality, measurement and measure, and concepts relating to data collection. Less frequently, concepts related to balance, blockmodels or equivalence appear. Sometimes, special issues are responsible for the appearance of tightly connected clusters of concepts, e.g., in the case of the special sections on personal networks and exponential random graphs in 2007 or the special issue on network analysis of infectious diseases in 1995. Due to the low frequency at which topics reappear, obvious title words such as 'network' and 'social' are indispensable for the continuity throughout this animation.

When the same animation is shown against the background of all title words included—at http://www.leydesdorff.net/journals/socnetw/ti/ti_aggr.htm—the point made above about coauthorship relations can be made with even more saliency: the title words in each year are specific instantiations of a semantic structure in the relevant repertoire (Figures 7a and 7b). This repertoire does also contract: the words at the margin are drawn into the center when used, and this reshapes the network to some extent. However, after a while sufficient structure is available in the background to provide a mental map against which the variation can be recognized as instantiations of a vocabulary.

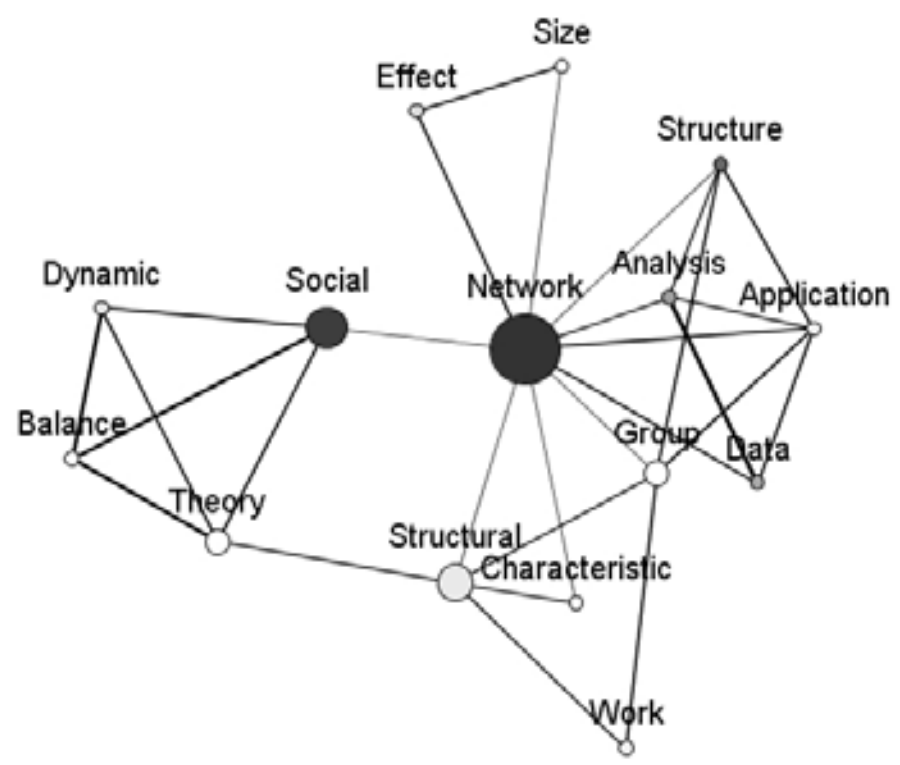

**Figure 7a:** Semantic map of title words in *Social Networks* 2003.

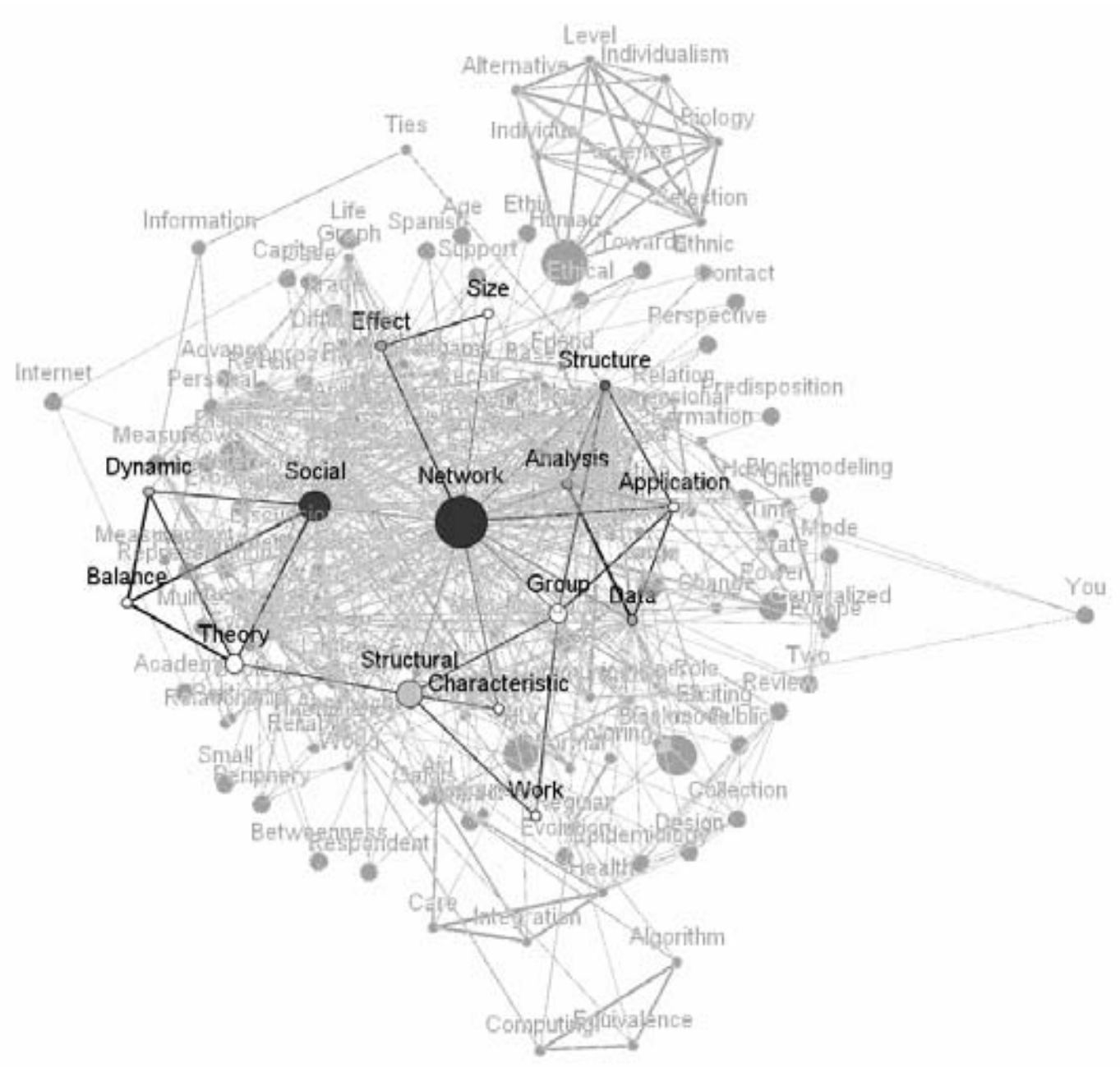

**Figure 7b**: The semantic map for 2003 as an instantiation of the vocabulary.

In other words, in each year topical issues addressed in journal articles draw on the vocabulary of a discourse which is shaped and reproduced at the level of the specialty. The title words in the publications of each year provide a specific selection from this larger repertoire.

*d. Title words from Google Scholar*

Would this volatility in the representations not be a consequence of the relatively small samples in each year? As shown in Figure 1 above, the number of publications per year is approximately twenty until 2003. Thereafter, it climbs to 45 in 2007, but this is still a limited set. Adding other journals (e.g., the *Journal of Mathematical Sociology*) to the set would not really solve this problem—because twice a small set remains a small set—but might even lead to more heterogeneity in the sets. Another option would be to include words in the abstracts of the articles under study, but this contextualization would also add a systematic source of variance (Leydesdorff, 1989).

Large sets on specific topics in scholarly publications can nowadays easily be generated at the Internet, for example, by using Google Scholar. Using the search string 'intitle:"social network" OR intitle:"social networks",' 6,071 titles were harvested on March 25, 2008 from Google Scholar. The distribution of these titles over the years is provided in Figure 8.

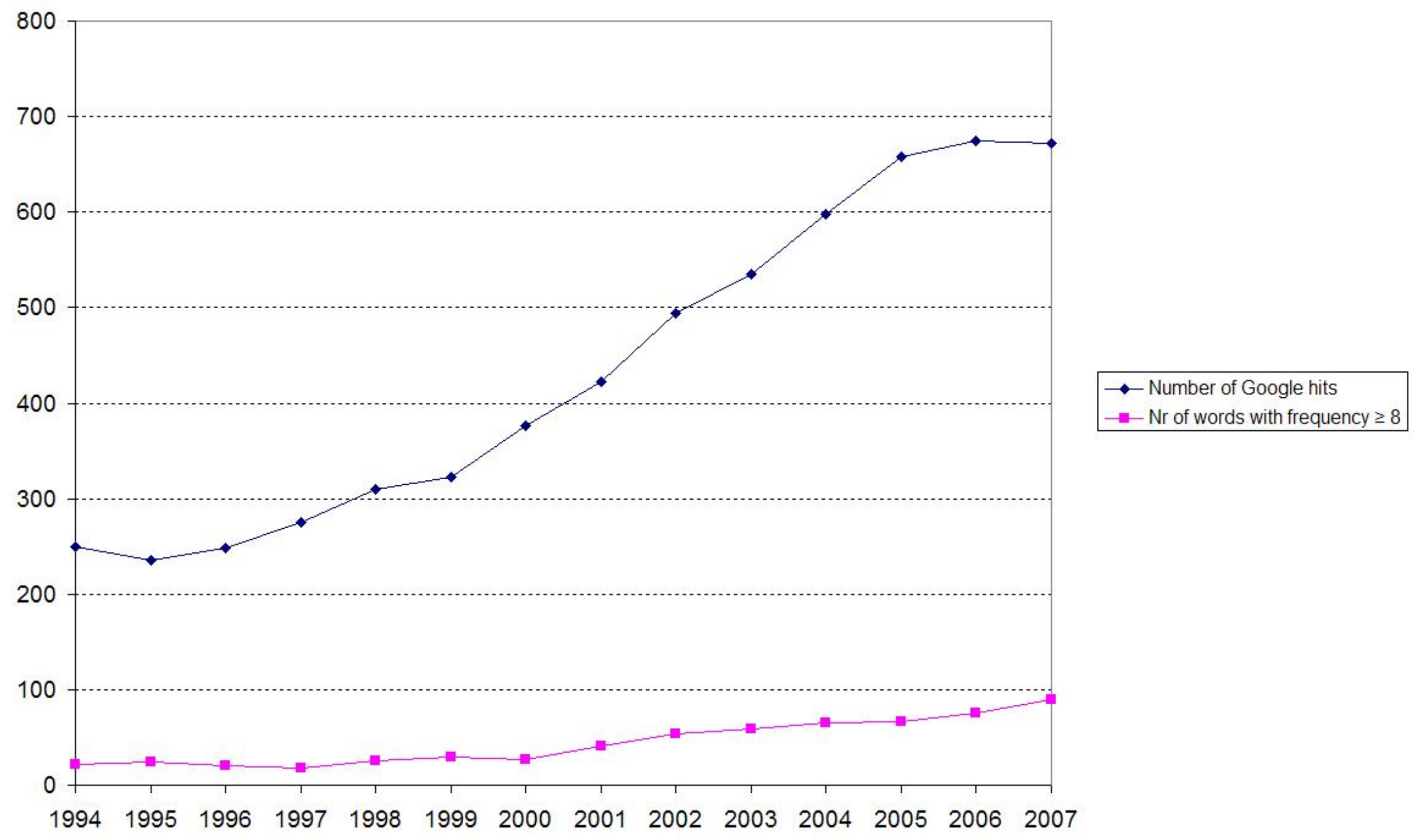

**Figure 8:** Number of hits with the title words "Social Network" or "Social Networks" for various years using Google Scholar (25 March 2008).

Indeed, this set is an order of magnitude larger than the one based on the journal *Social Networks.* After correction for the stopwords, 5,632 words occur 46,692 times of which 2,676 appear only once, and hence 2,956 more than once. After removing 'social' and 'network'—because these were the connecting words which therefore would dominate all visualizations—we used the 172 remaining words which occurred eight or more times in any single year. The resulting animation is brought online at http://www.leydesdorff.net/socnetw/googlescholar/index.htm.

The semantic structure in which the networks for each year appear, is more constant than in the previous animation. In the later years, that is, when more words are involved (see the second line in Figure 8), the occupation of this structure in each year is increasingly dense, and the instantiations therefore show more of the underlying structure. (We used $\omega$ = 4 for the stability factor (Eq. 5) in this case in order to stabilize the mental map; otherwise, the instantiations remained too volatile from year to year.) Like in the previous case, each event changes the structure, but the change in structure is less than the change in the variation.

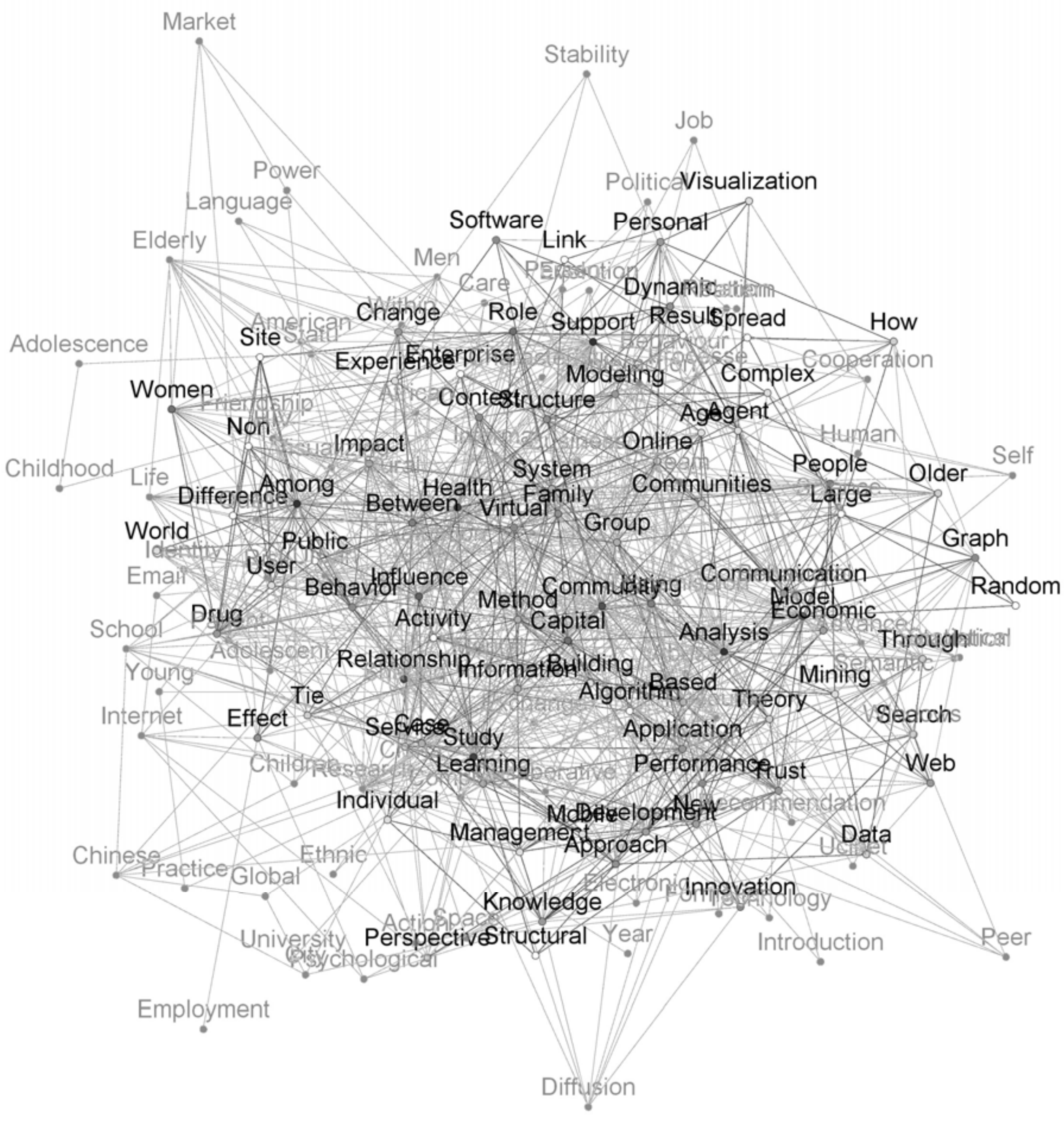

**Figure 9**: The network of 87 words instantiated in 2007 superimposed on the aggregated network of 172 words for the period 1994-2007. (N = 6071; cosine ≥ 0.1; $\omega$ = 4.)

Figure 9 shows the instantiations of 2007 against the structural background of the repertoire. It is worth noting that more technical concepts such as centrality, measure, equivalence, are hardly found. Instead, we find many title words relating to social networks such as social capital, as in having friends, relatives, or supporters. The repertoire is dominated by concepts that refer to less-privileged social groups, such as minorities, women, patients, and the elderly. In our opinion, scientific communities studying social cohesion rather than the vocabulary of scholars developing techniques for social network analysis are retrieved. However, these communities are not separated from the more methodologically oriented group around *Social Networks*.

In 2003, the software program UCINET developed by authors who are central in the *Social Networks* co-authorship network, surfaces in this repertoire. At first, technical matters seem mainly related to dynamic analysis, while later in the period data mining and visualization appear in this domain. The words 'method' and especially 'model' move towards the center in more recent years, suggesting that methodological issues pioneered in *Social Networks* gain increasing prominence among scholars investigating social networks. The titles of most publications, however, focus on substantive issues rather than technical ones.

It is interesting to backtrack the trajectories of the concepts that are most central in the 2007 network (Figure 10). Both the words 'group' and 'method' are central in this year. They appeared for the first time in the network in 2001, but at the margin, and immediately disappeared in the year thereafter. However, they were reintroduced in the network's center in 2004 and 2005, respectively, and remained there since. In contrast, the concepts 'community,' 'capital,' and 'system' progressed gradually towards the center of the network after their introduction in 1995, 1998, and 2002, respectively. These words seem to have acquired new meanings in the course of time.

'Capital,' for example, was first linked to 'political,' but thereafter it was tied to 'relation' and 'community,' suggesting that it is used increasingly in the context of interpersonal relations. In later years, it is linked to 'job' presumably due to Mark Granovetter's (1995 [1974]) study entitled *Getting a Job. A Study of Contacts and Careers*. In the later period, 'capital' is also connected to 'business' and 'economic.' In summary, 'capital' seems to have become a central catchword for the application of social network analysis across the social sciences.

The words 'community' and 'system' show similar albeit less pronounced developments. 'System' is first linked to 'economic' and 'analysis' in the periphery, but it gradually expands its connections to other concepts, notably 'method.' 'Community,' originally linked to 'health,' gets links to very diverse concepts such as 'capital,' 'ethnic,' 'building,' 'online,' 'people,' and 'group.' In this case, the dynamic visualizations show both a rather stable structure consisting of the main fields of application of social network analysis, and the dynamic rise of methodological reflection and theoretical concepts such as 'capital' and 'community.'

## Conclusions and discussion

A scientific journal can be considered as a niche of scientific communication entailing a specific set of authors, words, and cited references. Each journal is part of a larger network system of scientific communication including the journals of very different disciplines (Bradford, 1934; Garfield, 1972). The network of aggregated citation relations among journals relations can be considered as a next-order system (Leydesdorff, 1995). This network system provides a frame for each single journal such as *Social Networks*. The position of a journal in this reference system may change relationally without changing the structural (e.g., disciplinary) dimensions of the system. From this next-order perspective, the citation relations of a journal in a specific year provide the variation, while structures are reproduced over the years.

We have seen above that the perception of *Social Networks* in terms of its being cited in these journal environments, varies over the years more than the focus of the references provided by the authors publishing in the journal. In the citing dimension, authors construct and reconstruct the identity of a journal, while in the cited dimension the archive is selectively reproduced. One could consider journals as searching agents on a landscape (Scharnhorst, 2001). However, we found the search agent to be more specifically reproduced than the landscape because other search agents (journals) move in and out of the citation impact environment of *Social Networks.* The prevailing citation behavior of authors publishing in this journal is community based.

Because the disciplinary dimensions of the next-order journal system are structural from the perspective of each journal, we used the vector space (that is, cosine-normalized matrices) for the respective animations in the "cited" and "citing" dimensions. This structural perspective enables us to organize the topology in terms of its latent dimensions. However, the construction of the discourse in the journal is carried by a community of authors who may or may not coauthor with one another from year to year. We used coauthoring as a relational indicator for the bottom-up construction of the network. This bottom-up construction is relational, but the constructed system thereafter contains a structure which feeds back as a selection mechanism (e.g., quality control) on new variation by positioning the agents. As the network gravitates towards a coherent structure, codification of the discourse and therefore more codified citation patterns can increasingly be expected.

For mapping the discourse we assumed the generation of a semantic structure (Leydesdorff & Hellsten, 2005) and therefore adopted again the structuralist perspective of using cosine-normalized word-document matrices. The animations show how different domains in this semantic structure are instantiated in the various years and how these events change the structure by introducing new relations. When the semantic domain was enlarged by delineating it using Google Scholar, the semantic structure became more stable. However, we seem to have measured a different set of scientific communities using this latter database, notably one with more focus on substance than methods. In general, the semantic networks for each specific instance (year) can be considered as a retention mechanism: as words are used, they are repositioned and the network is reconstructed. The position of some words changes more than others while in each instantiation a large group of words also remains latent.

Are changes in the composition and structure of the impact and reference environments of *Social Networks*, its author community, and its semantic map to be considered as structural or rather fluctuations? The animations serve us primarily to infer hypotheses about latent dimensions. For example, they may help us to designate changes in the structures which could be tested in terms of static factor analyses in each of the years. Is there reason to assume a change in the number of principal components to be included in the factor analysis? Our analysis seems to indicate that the disciplinary basis and function of the journal has not changed; changes can be considered as fluctuation within a prevailing pattern. However, the use of specific layout and visualization techniques makes changes visible. A comparison with other dynamic visualization tools may shed light on the nature of these changes. To what extent are they artifacts of our methods?

Skye Bender-deMoll was so kind as to feed our journal matrix of *Social Networks* into *SoNIA*. The results were brought online as a QuickTime movie at http://skyeome.net/movies/leySnJournalII.mov (mirrored at http://www.leydesdorff.net/socnetw/sonia/sonia.mov). In this case, the animation itself explains the mechanism: all journals are initially placed on a circle and drawn into the network in the years that they are connected. When the journals leave the network in another year, they return to an open position on the enveloping circle. The focus of *SoNIA* is on the events in the middle of the circle.

Using *SoNIA*, we tried to generate an animation in accordance with the animation shown above, at http://www.leydesdorff.net/socnetw/sonia/index.htm. One recognizes the same structure as in the corresponding animation using *Visone* (at http://www.leydesdorff.net/journals/socnetw/index.htm). However, the drawing in of new journals from the (latent) circle disturbs the mental map in the case of the animation using *SoNIA.* The massive movement of appearing and disappearing journals prohibits a focus on the dynamics within the set of journals that remain in the layout.

The dynamic layout of *Visone* introduces new nodes (in this case, journals) in an anticipatory mode. The new nodes are introduced in the year before because the positions in the year ($t + 1$) are anticipatorily included in the computation of the year $t$, and so too are the positions in the year ($t - 1$). Thus, the new nodes do not come from an outer environment, but are generated within their context. Similarly, nodes can disappear locally.

We also used the same matrices as input into *PajekToSVGAnim.exe*, and uploaded the resulting animation at http://www.leydesdorff.net/socnetw/index.htm. The animation results generated using this routine are not essentially different from the ones generated using *Visone*, but for reasons specified above we submit that our results improve on these animations because they are not based on linear interpolations in a design which uses comparative statics. The differences might have been clearer if the trajectories would happen to have diverged more significantly between these two animations.

Nevertheless, if one focuses on the position of the journal *Social Networks* in the animations, it is clear that the *Visone* approach produces much more stable results while still conveying the same information. In the animation produced with *PajekToSVGAnim.exe*, *Social Networks* moves from one side of the sociological cluster to the other from 1997 to 1998, and it moves from one side to the other and immediately back again between 2000 and 2002. Constraining movement to the preceding and following year is very effective in avoiding this.

There is a fundamental difference in providing stability by using an initial layout (either common for all times or that of the previous moment) followed by an iterative layout procedure, and our approach that includes stability in the optimization. The outcome of the first approach can be stable or not; this depends on the relation to the layout at a previous point in time and the iterative procedure itself. The dynamic approach of *Visone* searches algorithmically for stability over time by considering the time axis as a third dimension of an array of matrices (networks) in which stress can be minimized. This approach allows for the extension to more than a single year in the future or the past, and can thus perhaps be made useful for our interest in the modeling of intentional systems which communicate meaning—meaning is provided from the perspective of hindsight!—in addition to information which is processed along the arrow of time (Dubois, 1998; Leydesdorff, 2008b, 2009; Luhmann, 1984; White, 1992).

**Acknowledgement**

The authors acknowledge Skye Bender-deMoll and James Moody for their assistance in working with *SoNIA*.